# A Network Topology for Composable Infrastructures

Opeyemi O. Ajibola, Taisir E. H. El-Gorashi, and Jaafar M. H. Elmirghani

*School of Electronic and Electrical Engineering, University of Leeds, LS2 9JT, United Kingdom*

**ABSTRACT**
This paper proposes a passive optical backplane as a new network topology for composable computing infrastructures. The topology provides a high capacity, low-latency and flexible fabric that interconnects disaggregated resource components. The network topology is dedicated to inter-resource communication between composed logical hosts to ensure effective performance. We formulated a mixed integer linear programming (MILP) model that dynamically creates logical networks to support intra logical host communication over the physical network topology. The MILP performs energy efficient logical network instantiation given each application's resource demand. The topology can achieve 1Tbps capacity per resource node given appropriate wavelength transmission data rate and the right number of wavelengths per node.
**Keywords**: silicon photonics, disaggregated datacentre, software defined network, composable infrastructure, software defined infrastructure, optical backplane.

## 1. INTRODUCTION

Digital transformation is driving ubiquitous demand for computation by end-users, enterprises and governments. Cloud computing which offers on-demand access to computation over the Internet has become the de facto means for computing capacity consumption because of its numerous benefits. Between 2010 and 2018, the volume of compute instances in datacentres (DCs) increased by over 500% [1]. As a result, the number and size of cloud DCs has increased in the last decade. More recently, the concept of edge computing is being adopted to augment the perceived shortcomings of cloud computing by bringing computation closer to end users at the edge of the network. This trend is particularly driven by the big data produced by new applications and services at the edge of the network [2] - [5]. Therefore, edge computing is increasingly needed to support the emergence of new applications and technologies such as Internet of Things and vehicle to everything communication while also offering on-demand consumption of computing capacity at the edge of the network. Further adoption of the edge computing concept is expected as more 5G infrastructure is deployed [6] – [8] and as succeeding 6G technologies emerge.

Amidst this growth in the use of computation capacities and the variety of ways to consume their underlying resources, increased attention is being focused on the sustainability of infrastructures that provide computation and on communication networks [9]-[14]. Attention has thus been paid to the energy efficiency of data centres networks [9] – [11], content distribution networks [12] – [16] and the core networks that support and interconnect data centres [17] – [24]. Traditionally, cloud computing services are supported by large clusters of high-end servers which are situated in remote hyper-scale DCs. Compared to 1% in 2010, a forecast predicts such DCs will account for 3-13% of global electricity consumption in 2030 [25]. Poor utilization efficiency due to resource fragmentation and stranding in traditional servers is partly responsible for high power consumption in today's DCs. Since edge processing nodes are relatively less energy efficient compared to hyper-scale DCs, the adoption of similar server architecture in distributed edge nodes of the emerging edge computing paradigm will further increase the high electricity consumption of the global ICT sector. The use of virtualization and software defined technologies while introducing improvements, has failed to completely address this problem in traditional DCs. In recent times, the concept of composable infrastructure has been proposed to enable greater efficiencies, flexibility and agility in computing infrastructures of all sizes [26].

Composable infrastructure leverages on the disaggregation of traditional server's intrinsic resources into physical or logical pools of homogeneous resources. A physical pool is a node which comprises of homogenous resources as shown in Rack 2 of Fig. 1a while a logical pool of homogenous resources is created on-demand from multiple homogenous or heterogeneous nodes. These pools are orchestrated on-demand over an appropriate network to create logical hosts that support end users' applications. The concept of resource disaggregation addresses the problems of resource fragmentation and stranding which is responsible for poor resource utilization in traditional servers. Resource disaggregation can be implemented physically or logically at different scales i.e. rack-scale, pod-scale or DC-scale [27]. The composition of logical hosts from disaggregated resources adopts virtualization technologies and other software-oriented techniques to abstract the control plane of physical resources from their data plane and for control, orchestration and monitoring. The availability of a suitable network interconnect between disaggregated resource components complements disaggregation and software-oriented techniques in composable infrastructures. However, this is slightly challenging as this interconnect must implement functions of the low latency and high bandwidth links associated with the intrinsic backplane of traditional servers at higher tiers of the DC network fabric.

Adoption of optical communication provides a practical solution to satisfy communication requirements in composable infrastructure, because it mitigates or avoids some known problems of electrical communication. However, sole use of optical communication infrastructure is not feasible because computation is performed in the electrical domain and optical buffering capabilities are limited. Therefore, hybrid opto-electronic communication networks enabled by silicon photonics technologies are widely expected to support the implementation of high-capacity, low latency and flexible networks to be used in composable infrastructures [27], [28]. Notwithstanding, significant maturity of silicon photonic technologies is required to enable practical and cost-effective extension of optics to both on-board and on-package levels of next-generation computing infrastructure.

In recent times, academic and industry research communities have proposed electrical, optical and hybrid network topologies for the different scales of composable infrastructures. In [29] and [30] Huawei and Intel respectively used electrical switches to interconnect different nodes present in the racks of their proposed composable infrastructure. The authors of [31] proposed a hybrid network topology for pod-scale composable infrastructure. Different variants of an all-optical network topology were proposed for composable infrastructure by authors of [32] – [34]. Authors in [35], [36] adopted two-tiers of optical switches in each rack of the dRedbox project. In this paper, we propose a hybrid network topology for rack-scale composable infrastructure. This novel topology maintains all-to-all direct connectivity between co-rack nodes while minimizing the number of interfaces required per node. This is achieved via the adoption of wavelength division multiplexing (WDM) techniques and the use of passive optical components. This paper describes the novel network topology and evaluates its performance and scalability by investigating its ability to setup suitable virtual links on-demand to support logical host instantiation in composable infrastructures. A mixed integer linear programming (MILP) model is formulated to conduct these studies.

## 2. A HYBRID NETWORK TOPOLOGY FOR COMPOSABLE INFRASTRUCTURE

The proposed hybrid network topology for composable infrastructure leverages on silicon photonics, passive optical components, WDM, and bidirectional communication over optical fibre to minimize the number of interfaces required at each node in a rack. Furthermore, a broadcast and select mechanism is employed to facilitate inter-node communication in the proposed optical backplane and to avoid contention during communication. As shown in Fig. 1, we focus on the design of a network topology for rack-scale composable infrastructure because it is easier to construct [37]. Moreover, our previous work in [38] showed the ability of rack-scale composable infrastructure to achieve similar efficiency as pod-scale composable infrastructure if resource-component allocation to racks ensures that resources are available in the appropriate amount and type. In addition, rack-scale composable infrastructure provides a modular design which can be replicated to implement a large-scale deployment in hyper-scale DCs or a small deployment at the edge of the network. The network topology forms an optical backplane to interconnect nodes (pool of resources) within a rack-scale composable infrastructure. An optical backplane may also be adopted to interconnect resources within each node; however, the scope of the proposed network topology gives little attention to such intra-node backplane.

Each node in the rack-scale composable infrastructure has a node controller hub (NCH) which replaces the platform controller hub of traditional servers. As shown in Fig. 1, all resource-components in a node are connected to the NCH. These co-node resource components may also maintain direct connectivity to one another via the node's on-board fabric to reduce the workload on the NCH and to ensure path diversity within the node. The NCH is a network element which performs network related computation in the proposed network topology. It may be implemented on a specialised ASIC in commercial deployment and by a FGPA in experimental scenarios. The NCH perform functions such as end to end virtual network setup for inter-node communication via the assignment of wavelengths for direct node to node communication; the multiplexing of data onto and the de-multiplexing of data from assigned inter-node wavelengths; and it may also serve as an intermediate node on an indirect multi-hop path between two nodes. The NCH is integrated with each node's interfaces to ensure coordination of wavelength selection for hop to hop communication over bidirectional optical links. These optical links form the passive optical backplane in each rack.

To promote wavelength reuse in each rack and to minimize the number of unique wavelengths required per modular rack, each node has two interfaces. Adoption of two interfaces per node also enables path diversity which improves the resilience and capacity of the network topology. Each interface comprises of an array of optical transceivers that transmit and receive a set of pre-defined wavelengths. The sets of transmitting and receiving wavelengths on each optical fibre link must be mutually exclusive to minimize the impact of crosstalk noise in the system. Wavelengths used for transmission by an interface are used for receiving by the other interface and vice versa. This enables greater economies of scale in the use of wavelengths as the node can use all the wavelength for transmitting and receiving. Bi-directional communication on each fibre optic link is enabled by the use of optical circulators between the fibre links and the splitters/combiners. This helps to reduce the size of the rack backplane by half relative to the use of unidirectional links or communication paths. The same type of interface is used in all nodes in each rack to further leverage on the benefits of economies of scale and to enable easy replication. Integration of the NCH element and node interfaces may be implemented as a co-packaged device with optical I/O by leveraging on silicon photonics technologies.

In the transmitting direction, the wavelengths transmitted by both interfaces of each node flow through optical circulators to the multiplexer, which combines all transmitted wavelengths of each node. Wavelength collision is avoided using parallel paths. The multiplexer is connected to a splitter which broadcasts the transmitted wavelengths to all nodes within the same rack and to the top of rack (TOR) switch. The TOR switch supports intra-rack and inter-rack communications. A low latency electrical switch such as the switch proposed by the Gen-Z consortium [39] may be adopted as the TOR switch. A node in a rack is connected to all other co-rack nodes and to the TOR switch via bidirectional optical links.

In the receiving direction, a combiner receives all transmitted wavelengths from other co-rack nodes and the TOR switch and forwards the received wavelengths to a de-multiplexer. The de-multiplexer separates and forwards each received wavelength to the corresponding circulator that leads to the receiving interface. At the interface, each transceiver receives its associated wavelength and forwards the received data to the NCH. The NCH de-multiplexes the received data and forwards to the appropriate resource-component if it is in the destination node. Otherwise, the NCH forwards the received data to the corresponding interface linked to the next hop on the multi-hop communication path by selecting an appropriate wavelength(s). Each rack transmit-receive wavelength pair is limited to that rack. Hence, the same transmit-receive wavelength pairs can be reused in other racks. Splitters (combiners) support the broadcast and select mechanism adopted for the network topology by broadcasting (receiving) the total egress (ingress) traffic of each node onto (from) the optical backplane. Hence, splitters/combiners reduce the number of interfaces required at each node to form a full mesh optical backplane in a rack.

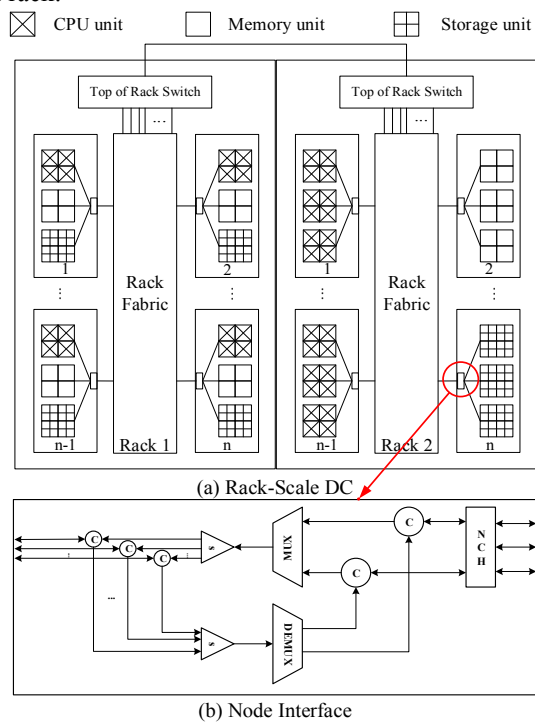

*Figure 1. Proposed network topology.*

*Table 1. Model input parameters.*

| Parameters | Value |
|---|---|
| Node controller hub energy per bit | 1.4pJ/b [40] |
| Top of rack switch energy per bit | 0.028pJ/b |
| Top of rack switch idle power | 312W |
| On-board energy per bit | 0.1pJ/b |
| CPU resource demand | 0.9/1.8/2.7GHz |
| Memory resource demand | 3.6/7.2/10.8/26/32GB |
| Storage resource demand | 80/160/240GB |
| CPU capacity, peak power and dynamic power range | 3.6GHz, 130W and 30% |
| Memory capacity, peak power and dynamic power range | 32GB, 11.85W and 30% |
| Storage capacity, peak power and dynamic power range | 320GB, 6.19W and 30% |
| CPU-Memory communication data rate per application | 300-800Gbps |
| CPU-Disk communication data rate per application | 5 – 128Gbps |
| Single wavelength transmission data rate | 50Gbps or 100Gbps |

*Table 2. Maximum capacity per node.*

| Wavelengths per node | 2λ | 4λ | 6λ | 8λ | 10λ |
|---|---|---|---|---|---|
| Capacity (Tbps) per node at 50Gbps single-λ data rate. | 0.1 | 0.2 | 0.3 | 0.4 | 0.5 |
| Capacity (Tbps) per node at 100Gbps single-λ data rate. | 0.2 | 0.4 | 0.6 | 0.8 | 1 |

At each node, the NCH performs wavelength selection based on global knowledge of the selection made in other nodes. Hence, all NCH in each node must be centrally controlled and orchestrated to ensure optimal wavelength utilization and the ability to operate the system at maximum capacity. The broadcast and select mechanism deployed in the network topology implies that each node only accepts a predefined list of wavelengths at its interface at a given moment while other received wavelengths are discarded.

## 3. TOPOLOGY PERFORMANCE EVALUATION

### 3.1 Model Description and Evaluation Scenarios

To evaluate the performance of the proposed physical network topology in supporting on-demand instantiation of virtual networks used by applications, we formulate a MILP model. The model performs application resource demand placement and routing and wavelength assignment in a rack-scale composable infrastructure of logically disaggregated nodes that uses the proposed network topology. The MILP model ensures the satisfaction of both network and compute related constraints. The objective of the MILP model is to minimize the total network power consumption (TNPC), total compute power consumption (TCPC), the total number of rejected applications (TRA) and the total number of active wavelength (TAW) in the proposed topology as illustrated in Eq. (1). We minimize the total number of active wavelengths to conserve inactive wavelengths for future use or to improve the system's resilience.

$$Minimize: \alpha_1 \cdot TNPC + \alpha_2 \cdot TCPC + \alpha_3 \cdot TRA + \alpha_4 \cdot TAW \qquad (1)$$

where $\alpha_1$ is the cost associated with TNPC; $\alpha_2$ is the cost associated with TCPC; $\alpha_3$ is the cost associated with TRA; and $\alpha_4$ is the cost associated with TAW in the proposed fabric. Under all evaluation scenarios, $\alpha_2 = \alpha_4 = 1$ while $\alpha_3$ is set to a very large number to emphasize the high cost of application rejection. Two different values are adopted for $\alpha_1$ during evaluation, $\alpha_1 = 1$ and $\alpha_1 << 1$ in the first (I) and second (II) evaluation scenarios respectively. Adoption of a very low value for $\alpha_1$ significantly reduces the impact of TNPC in the formulated model. We consider two data rates i.e. 50Gbps and 100Gbps for single wavelength transmission in the network topology. The number of transmit-receive wavelength pair per interface is also varied between 1-3 wavelength pairs per node. The power consumption of computing (CPU, memory and storage) resources are estimated by considering the idle power consumption and the load proportionate power consumption over each resource's dynamic power range as illustrated in Table 1. Likewise, the power consumption of the adopted electronic TOR switch comprises of both idle and load (traffic) proportionate portions. We adopt a load proportional power consumption profile for the NCH element in each node. We conservatively assume that each NCH has the same energy per bit values as a 10Gbps commodity off-the-shelf offload network interface card that has peak power of 14W [40].

We consider a small rack-scale composable infrastructure with 9 heterogeneous nodes and an electronic TOR switch. Each node consists of a CPU, a RAM and an HDD along with the NCH element which is integrated with two bespoke interfaces as described earlier. We adopt 15 input applications with a mix of compute resource demand intensity to evaluate the performance of the proposed network topology. Each input application has inter-resource (CPU-memory and CPU-disk) communication requirements which are generated using uniform distribution over the ranges given in Table 1.

### 3.2 Results and Discussions

Whilst a single wavelength data rate is 50Gbps, an application is rejected under both scenarios I and II when each node in the rack-scale composable infrastructure can only transmit and receive 2 wavelengths as shown in Figure 2. This rejection is due to network bottlenecks which prevent the use of physically disaggregated resource components to support the rejected application. Compared to other scenarios where the data rate of single wavelength has increased or the number of the wavelengths per node has increased, this rejection is responsible for the relatively lower TCPC under scenarios I and II when 2 wavelengths are used as shown in Figure 3. For both scenarios I and II, an increase in the number of wavelengths per node or the data rate of single wavelength transmission prevents application rejection in the composable infrastructure as shown in Figure 2.

We observe that the use of resource disaggregation concept in the composable infrastructure is significantly restricted, i.e. an application is served using resources within the same node, under scenario I irrespective of the number of wavelength pairs available to each node in the rack or the data rate of single wavelength transmissions. Restricting the disaggregation helps to reduce the network power consumption by reducing the volume of traffic on the fabric of the composable infrastructure (due to inter-resource traffic). When each node has 2 wavelength pairs under scenario I, disaggregation is completely avoided in the rack-scale composable infrastructure. The composable infrastructure relies solely on virtualization to enable consolidation of applications which leads to optimal TCPC; hence, the NCH makes no contribution to the TNPC as shown in Figure 4. However, as the number of wavelength pairs per node increases, minimal implementation of disaggregation is adopted to avoid application rejection. Since the presence of inter-resource traffic on the rack backplane can lead to significant increase in TNPC, only applications with low inter-resource traffic are provisioned on disaggregated resource components to minimize TNPC. As a result, all resource components are active in scenario I as shown in Figure 5. However, the TNPC under scenario I is relatively lower compared to scenario II.

The high load proportional power consumption of the NCH element present in each node is responsible for a large portion of the TNPC as shown in Figure 4. Hence, it is a strong factor in the objective function of the formulated MILP model. Adoption of a very low value for $\alpha_1$ (i.e. $\alpha_1 << 1$) under Scenario II significantly reduces the impact of TNPC (and NCH power consumption) in the formulated model. Scenario II represents a situation where technological advancements have significantly improved the power consumption of the NCH and its integrated interfaces and other network components. A general trend observed under scenario II for different number of wavelengths per node is that there is an increase in the number of instances where applications are provisioned on disaggregated resource components. As a result, there is an increase in the number of inactive resource components in the composable rack-scale infrastructure as shown in Figure 5. Relative to scenario I, the TCPC of scenario II falls by almost 1% if all applications are provisioned (otherwise, 9% reduction in the TCPC is observed when an application was rejected under 2 wavelength per node scenario). This is achieved at the expense of increased TNPC relative to Scenario I as shown in Figure 4. However, as reported under Scenario I, attempts are also made to curtail the rise in TNPC under scenario II by ensuring that only applications with low inter-resource (CPU-memory and CPU-storage) traffic are provisioned on disaggregated resource components in the composable infrastructure.

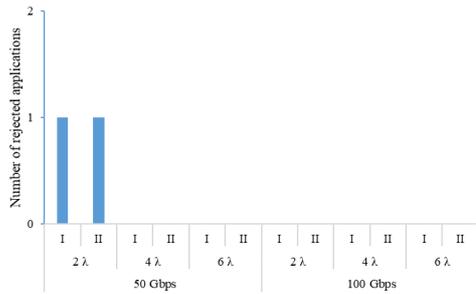

*Figure 2. Number of rejected applications.*

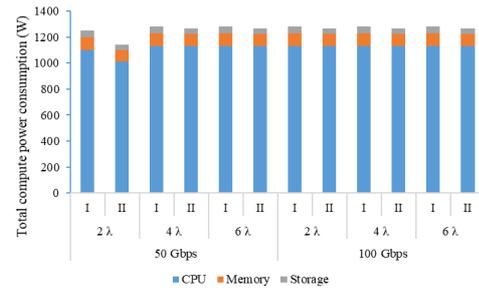

*Figure 3. Total compute power consumption.*

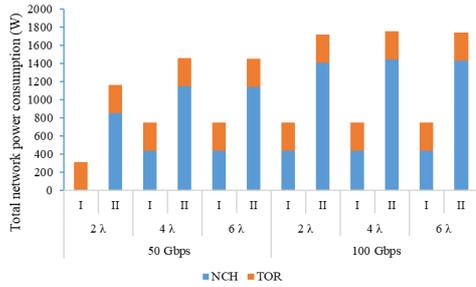

*Figure 4. Total network power consumption.*

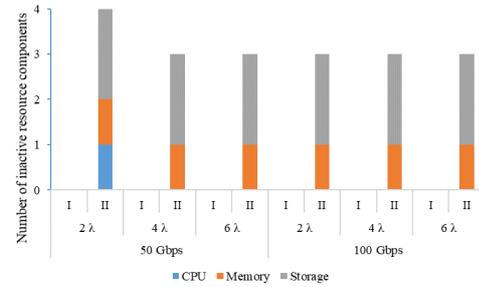

*Figure 5. Number of inactive resource components.*

As observed in Figures 3 and 5, increasing the data rate of a single wavelength transmission from 50Gbps to 100Gbps does not lead to reductions in the (minimum) TCPC. However, Table 2 shows that the maximum capacity per node in the proposed network topology is directly proportional to both the number of wavelength pairs supported by the node and the single wavelength data rate. Although, there are practical limits to the number of transceivers that can be integrated with the NCH element in the proposed network topology, present and future increase in single wavelength data rates promises to enable higher capacity. For example, Table 2 shows that 1Tbps capacity per node can be achieved using the proposed network topology for rack-scale composable infrastructure if each node can transmit 10 disjoint wavelengths using 100Gbps single wavelength rate. This can be very beneficial in scenarios where the deployment of cloud native applications such as micro-services in composable infrastructures leads to significant increase in the volume of ingress and egress traffic per node. Furthermore, increase in single wavelength data rates can also promote the implementation of time and wavelength division multiplexing (TWDM) in the proposed network topology to enable greater flexibility and granularity.

## 4. CONCLUSIONS

In this paper, a new network topology for a rack-scale composable computing infrastructure was introduced along with the description of its operating principles. Results showed that implementation of logical disaggregation is restricted when network power consumption is high, and that applications rejection can occur due to network bottlenecks as a result of low single wavelength transmission data rate and small number of transmission wavelengths. On the other hand, implementation of virtual disaggregation increases in composable infrastructure when network capacity suffices to enable reduced compute power consumption. To ensure minimal network power consumption due to the exchange of inter-resource traffic over the proposed optical backplane, higher priority is given to applications with low inter-resource traffic when selecting the applications to be provisioned on disaggregated resource components. Higher single wavelength data rates will increase the capacity of the proposed network topology.

## ACKNOWLEDGEMENTS

The authors would like to acknowledge funding from the Engineering and Physical Sciences Research Council (EPSRC), INTERNET (EP/H040536/1), STAR (EP/K016873/1) and TOWS (EP/S016570/1) projects. The first author would like to acknowledge the support of the Petroleum Technology Trust Fund (PTDF), Nigeria, for the Scholarship awarded to fund his PhD. All data are provided in full in the results section of this paper.